\documentclass[aip,reprint,amsmath,amssymb,author-numerical]{revtex4-1}

\usepackage{graphicx}% Include figure files
\usepackage{dcolumn}% Align table columns on decimal point
\usepackage{bm}% bold math
\usepackage[utf8]{inputenc}
\usepackage[T1]{fontenc}
\usepackage{mathptmx}
\usepackage{etoolbox}

\makeatletter
\def\@email#1#2{%
 \endgroup
 \patchcmd{\titleblock@produce}
  {\frontmatter@RRAPformat}
  {\frontmatter@RRAPformat{\produce@RRAP{*#1\href{mailto:#2}{#2}}}\frontmatter@RRAPformat}
  {}{}
}%
\makeatother
\begin{document}

\title[Accurate MP2 extrapolation from small basis set]{Accurate extrapolation of MP2 correlation energies for small correlation-consistent core-valence basis sets}

\author{E. Fabiano}
\affiliation{Center for Biomolecular Nanotechnologies, Istituto Italiano di
Tecnologia, Via Barsanti 14, 73010 Arnesano (LE), Italy}
\affiliation{Institute for Microelectronics and Microsystems (CNR-IMM), Via
Monteroni, Campus Unisalento, 73100 Lecce, Italy} 
\email{eduardo.fabiano@cnr.it}
\author{F. Della Sala}%
\affiliation{Center for Biomolecular Nanotechnologies, Istituto Italiano di
Tecnologia, Via Barsanti 14, 73010 Arnesano (LE), Italy}
\affiliation{Institute for Microelectronics and Microsystems (CNR-IMM), Via
Monteroni, Campus Unisalento, 73100 Lecce, Italy}

\date{\today}

\begin{abstract}
The Atom-Calibrated Basis-set Extrapolation (ACBE) method is introduced as a robust approach for extrapolating MP2 correlation energies from small basis sets. Unlike conventional extrapolation techniques, ACBE incorporates system- and environment-specific parameters to enhance predictive accuracy, effectively mitigating errors associated with finite basis sets. Evaluated using the aug-cc-pwCV$n$Z basis set family across a diverse set of molecular systems, including first- and second-row species, ACBE consistently delivers reliable energy estimates, even when double- and triple-zeta basis sets are employed. These results highlight the computational efficiency of the method, making it a promising option for large MP2 studies.  
\end{abstract}

\maketitle

\section{Introduction}
M\o{}ller-Plesset second-order perturbation theory (MP2) \cite{mp2} holds significant relevance in quantum chemistry and electronic structure theory as a foundational method for incorporating electron correlation into molecular calculations \cite{cremer11,fink16}. As an efficient method for capturing dynamic correlation effects beyond Hartree-Fock (HF), MP2 is particularly valuable for studying molecular properties, potential energy surfaces, and chemical reaction mechanisms with reasonable accuracy \cite{riley07,riley12,su19,loipersberger21}. Moreover, MP2 correlation energy is widely used in several high-level electronic structure approaches such as double-hybrid density functionals \cite{goerigk14} and M{\o}ller-Plesset adiabatic-connection methods \cite{seidl18,daas21,hfac24}.

Nevertheless, one of the primary challenges associated with MP2 is its slow convergence with respect to the size of the basis set \cite{varandas_rev18,helgaker_book}. This slow convergence arises mainly because the finite basis sets fail to fully represent the nuclear cusp as well as the continuum of virtual orbitals, which are crucial for MP2 electron correlation \cite{varandas_rev18,helgaker2001}. Moreover, small- or medium-sized basis sets often fail to accurately describe long-range correlation effects, leading to systematic biases and erratic convergence behavior \cite{makerewicz04}.

Incomplete basis sets inevitably lead, therefore, to truncation errors, significantly impacting the accuracy of MP2 calculations. As the size of the basis set increases, the computed MP2 correlation energy asymptotically approaches the complete basis set (CBS) limit. However, reaching this limit typically requires very large basis sets, resulting in substantial computational costs. Consequently, basis set truncation error remains a persistent issue, particularly for properties highly sensitive to electron correlation, such as interaction energies and dispersion forces. Furthermore, the basis set superposition error (BSSE), arising from the artificial lowering of the total energy due to overlapping basis sets of interacting fragments, further exacerbates these challenges, especially in systems with weak interactions \cite{vidal19,jensen24}.

To address slow convergence and truncation error, numerous strategies and methodological advancements have been proposed \cite{varandas_rev18}. These include dual-basis approaches \cite{deng11,zhang13}, the development of explicitly correlated methods \cite{kong12,hattig12,tenno12}, such as MP2-F12 \cite{bachorz11}, which introduce explicitly terms dependent on the interelectronic distance into the wavefunction, and machine learning techniques \cite{holm23,speckhard25} that aim to predict the MP2 CBS-limit energy directly \cite{han21}, thereby circumventing the need for large basis set calculations.

Nevertheless, the most widely adopted strategy remains basis set extrapolation \cite{varandas_rev18,martin96,helgaker97,halkier98,truhlar98,varandas00,huh03,schwenke05,varandas07,bakowies07_1,bakowies07_2,schwenke12,feller13,martin18,lesiuk19}, which uses MP2 energies computed with a series of systematically larger basis sets to extrapolate to the CBS limit. Various two- or three-point extrapolation schemes have been developed based on the asymptotic behavior of basis set convergence. These approaches enhance accuracy, maintain the simplicity of MP2, and avoid the computational demands of extremely large basis sets.

Most basis set extrapolation techniques rely on specially designed basis sets characterized by a cardinal number $n$, which corresponds to expansions with maximum angular momentum $n-1$ \cite{varandas_rev18}. A notable example is the Dunning family of cc-pV$n$Z basis sets and related variants \cite{dunning89,woon1993,kendall1992,peterson2002}. 
These conventional approaches typically neglect core-valence contributions and achieve reliable accuracy only when triple-zeta (TZ) or larger basis sets are used as a starting point. Moreover, they often struggle to treat first- and second-row atomic species on an equal footing. This limitation arises because, particularly when core-valence effects are included and small basis sets are employed, different systems exhibit distinct convergence behaviors toward the CBS limit. Such variations are challenging to capture using global extrapolation methods.

Theoretical studies of correlation in two-electron atoms \cite{schwartz62,schmidt83} suggest that the dependence of correlation energy on the basis set can be approximately expressed as
\begin{equation}\label{eq1}
E_n = E_\infty + Af(n)\ ,
\end{equation}
where $E_n$ is the energy computed with a basis set of cardinal number $n$, $E_\infty$ represents the CBS limit, $A$ is a constant, and $f(n)$ is an attenuation function dependent on $n$.
When applied to two different basis sets, with cardinal numbers $n_1$ and $n_2$ (typically $n_2 = n_1+1$), the CBS correlation energy can be obtained as:
\begin{equation}\label{eq2}
E_\infty = E_{n_2} + \frac{f(n_2)}{f(n_1) - f(n_2)}\left(E_{n_2} - E_{n_1}\right).
\end{equation}

Finding the optimal form of $f(n)$ remains an open problem in CBS extrapolation. Many approaches are grounded in theoretical considerations. For instance, if $n$ is sufficiently large, higher-order contributions in $f(n)$ can be neglected, leading to the popular choice $f(n) = n^{-3}$, as proposed by Helgaker \cite{helgaker97} (denoted HELG). Another approach, introduced by Martin \cite{martin96}, accounts for different asymptotic behaviors of first-row atoms and hydrogen by offsetting $n$, resulting in $f(n) = (n+1/2)^{-3}$ (denoted MART).
More advanced methods include higher-order corrections, such as those by Varandas (VARA) \cite{varandas07}, which employ $f(n) = (n+\alpha)^{-3}\left[1+\tau_{53}(n+\alpha)^{-2}\right]$, with $\alpha = -3/8$ and a system-dependent constant $\tau_{53}$.

In practice, however, Eq. (\ref{eq1}) involves various approximations when applied to the general molecular context. Consequently, semi-empirical options are often employed, such as $f(n) = n^{-\beta}$, $f(n) = (n+\alpha)^{-3}$, or $f(n) = \exp(-\beta n)$, where $\alpha$ and $\beta$ are parameters fitted to training sets of atoms and molecules. For instance, Truhlar \cite{truhlar98} utilized the exponential ansatz to develop an extrapolation procedure for MP2 energies using small basis sets of double- and triple-zeta quality (denoted TRUH).

Finally, Bakowies \cite{bakowies07_1,bakowies07_2} explored the use of the exponential ansatz with either an optimized exponent or an optimized offset parameter. He observed that, particularly for small basis sets, the optimal parameter values (those that reproduce the CBS limit for a given system) vary significantly across different chemical species. To address this, he proposed system-dependent parameters derived by averaging the optimal values of the constituent atoms within a molecule. A similar methodology, emphasizing specific extrapolations, was adopted by Schwenke \cite{schwenke05}, who employed analytical rather than numerical scaling.

The method proposed by Bakowies is relatively straightforward and effectively captures the specific rate of convergence for different systems and basis sets, thereby improving the accuracy of basis set extrapolations from small basis sets. However, it has some limitations. First, it is only defined for first-row atoms, and due to its reliance on the count of $1s$, $2s$, and $2p$  orbitals, there is no straightforward way to extend it to elements beyond the first row. Additionally, the method does not account for the impact of intramolecular interactions on the extrapolation behavior of individual atoms. Finally, it lacks clarity on how to handle charged systems, further restricting its general applicability.

In this work we consider these issues and we propose a simple extrapolation approach, both for the DT and TQ case, that includes the specificity of different atomic species, can be naturally applied to all atoms and is automatically taking into account molecular charge and polarization.
This new approach is described in the next section and it is then optimized and tested for the molecules composed of first- and second-row atoms, using the family of aug-cc-pwCV$n$Z basis sets \cite{dunning89,woon1993,kendall1992,peterson2002}. With this choice of the reference basis set, the method also accounts for core-valence contributions in the MP2 correlation energy.

\section{Method}
The CBS energy is expressed using Eq. (\ref{eq2}), where the attenuation function is defined as
\begin{equation}\label{eq3} f(n) = n^{-\beta}. \end{equation}
To account for the system-specific convergence behavior, particularly the distinct trends observed for first- and second-row atoms, the parameter $\beta$ in Eq. (\ref{eq3}) is not treated as a global constant. Instead, it is computed as a weighted average of optimized atomic parameters:
\begin{equation}\label{eq4} \beta = \frac{\sum_i^{N_{at}}Z_i\beta_i^{opt}}{\sum_i^{N_{at}}Z_i},
\end{equation}
where the index $i$ runs over all the $N_{at}$ atoms, $Z_i$ denotes the atomic number, and $\beta_i^{opt}$ represents an optimized atomic parameter, discussed in detail below. While Eq. (\ref{eq4}) is structurally analogous to Eq. (35) in Ref. \cite{bakowies07_1}, our approach differs by employing a weighted average to account for the varying electronic contributions of different atomic species. Additionally, we introduce a novel definition of the optimized atomic parameters $\beta_i^{opt}$.

The atomic parameters in Eq. (\ref{eq4}) are not only element-specific but also incorporate the local electronic environment of each atom within the molecular framework. Specifically, they account for partial charge redistribution due to intramolecular interactions. To capture these effects, we define $\beta_i^{opt}$ using a quadratic dependence on the atomic partial charge:
\begin{equation}\label{eq5}
\beta_i^{opt} = \beta_i^{opt}(q_i) \equiv a_iq_i^2 + b_iq_i + c_i\ ,
\end{equation}
where $q_i$ is the partial charge of the $i$-th atom, determined in this work using Mulliken analysis. The coefficients $a_i$, $b_i$, and $c_i$ are species-specific and are calibrated such that, for each atomic species, the $\beta_i$ parameter reproduces the reference CBS correlation energy at three distinct charge states: $q=-1$ (anion), $q=0$ (neutral atom), and $q=1$ (cation).
For hydrogen and helium, the cationic case can not be considered. Consequently, for these elements, we adopt a linear dependence by setting $a_i=0$. Furthermore, for neutral hydrogen, the H$_2$ molecule is used as a reference system, because due to the homonuclear symmetry the $\beta_i$ value remains identical for isolated atoms and within homodimers.

Equations (\ref{eq3})–(\ref{eq5}) collectively define the atom-calibrated basis-set extrapolation (ACBE) method, a systematic and computationally efficient approach for estimating CBS energies with high accuracy. This method captures the system-specific convergence characteristics across different atomic species and is inherently adaptable to a broad range of molecular systems. Moreover, it seamlessly incorporates the effects of molecular charge and polarization, enabling a robust and transferable extrapolation scheme applicable to diverse chemical environments.

\subsection{Parameterization}
The only parameters requiring optimization are the atomic coefficients $a_i$, $b_i$, and $c_i$ in Eq. (\ref{eq5}). In this work, we have determined these coefficients for all elements up to argon, excluding lithium and beryllium, using the aug-cc-pwCV$n$Z family of basis sets \cite{dunning89,woon1993,kendall1992,peterson2002}.

Our procedure accounts for two extrapolation schemes: DT extrapolation, which utilizes results from the aug-cc-pwCVDZ and aug-cc-pwCVTZ basis sets, and TQ extrapolation, which employs results from the aug-cc-pwCVTZ and aug-cc-pwCVQZ basis sets.

As outlined above, for each atomic species (except hydrogen and helium), we considered three charge states: neutral, anionic, and cationic. For each case, Eq. (\ref{eq2}) was applied with $f(n)=n^{-\beta_{fit}}$, where $\beta_{fit}$ was determined to exactly reproduce the reference CBS energy within the corresponding extrapolation scheme.

The resulting $\beta_{fit}$ values were subsequently used to parameterize the quadratic expression in Eq. (\ref{eq5}), ensuring internal consistency and accuracy in the extrapolation process. The final optimized parameters are reported in Table \ref{table_abc}.

\begin{table}
\caption{\label{table_abc}Atomic-optimized coefficients $a_i$, $b_i$, and $c_i$ (see Eq. (\ref{eq5})) for all atoms and extrapolation schemes considered in this work.}
\begin{ruledtabular}
\begin{tabular}{lccc}
%\hline\hline
atom & $a_i$ & $b_i$ & $c_i$ \\
\hline
\multicolumn{4}{c}{DT-extrapolation}\\
H  & 0.00000 & 0.47869 & 2.70690 \\
He & 0.00000 &-0.21015 & 2.45396 \\
B  &-0.04209 &-0.04690 & 2.52323 \\
C  &-0.02326 &-0.03424 & 2.66750 \\
N  &-0.00021 &-0.03342 & 2.79236 \\
O  & 0.03717 & 0.15639 & 2.62673 \\
F  & 0.02069 & 0.10344 & 2.53995 \\
Ne & 0.16117 &-0.06538 & 2.48421 \\
Al &-0.00129 &-0.01062 & 1.47852 \\
Si &-0.00160 &-0.01613 & 1.53296 \\
P  &-0.00725 &-0.01416 & 1.58471 \\
S  &-0.00155 &-0.00723 & 1.60730 \\
Cl &-0.00160 &-0.00258 & 1.65496 \\
Ar & 0.04561 & 0.04309 & 1.69880 \\
\hline
\multicolumn{4}{c}{TQ-extrapolation}\\
H  & 0.00000 & -0.26456 &  2.68426 \\ 
He & 0.00000 & -0.18258 &  2.68710 \\ 
B  & 0.02509 &  0.12430 &  3.19450 \\ 
C  & 0.00940 &  0.08105 &  3.12643 \\ 
N  & 0.00014 &  0.06999 &  3.08717 \\ 
O  & 0.06459 &  0.16731 &  2.86032 \\ 
F  & 0.03898 &  0.06875 &  2.73865 \\ 
Ne & 0.27610 & -0.23172 &  2.66096 \\ 
Al &-0.00098 & -0.01141 &  1.68227 \\ 
Si &-0.00141 & -0.01347 &  1.84192 \\
P  & 0.00211 & -0.01187 &  2.06813 \\ 
S  & 0.00414 & -0.01395 &  2.10385 \\ 
Cl & 0.00297 & -0.02118 &  2.07976 \\ 
Ar & 0.06352 & -0.08843 &  2.10089 \\ 
%\hline\hline
\end{tabular}
\end{ruledtabular}
\end{table}

\section{Computational details}
All calculations were performed using the family of aug-cc-pwCV$n$Z basis sets \cite{dunning89,woon1993,kendall1992,peterson2002} and the TURBOMOLE program package \cite{turbomole,turbo_rev}. The resolution-of-identity (RI) approximation \cite{eichkorn95,weigend98} was employed throughout.

We tested the method on a set of 90 molecular systems: 

\textbf{First-row-molecules:} CO, HF, C$_2$H, CF$_4$, CH$_4$, CO$_2$, OF$_2$, H$_2$O, CNH, CHO, N$_2$O, NF$_3$, NO$_2$, C$_3$H$_4$, C$_2$H$_3$, CH$_3$O, CHF$_3$, COF$_2$, CH$_2$O, H$_2$O$_2$, N$_2$H$_4$, CH$_2$O$_2$, C$_2$H$_4$, NH$_3$, CH$_2$F$_2$, C$_2$H$_2$O, NO, NH, C$_2$H$_2$, C$_2$H$_3$F, C$_2$N$_2$, CH$_2$, N$_2$, F$_2$, CN, BF$_3$, CF$_2$, C$_2$H$_2$O$_2$, C$_4$H$_8$, CH$_4$-Ne, He-Ne, NH$_3$-H$_2$O, HF-HF, Ne$_2$, NH$_3$-NH$_3$, CH$_4$-CH$_4$, H$_2$O-H$_2$O$_2$, C$_2$H$_4$, NH$_3$, CH$_2$F$_2$, C$_2$H$_2$O, NO, NH, C$_2$H$_2$, C$_2$H$_3$F, C$_2$N$_2$, CH$_2$, N$_2$, F$_2$, CN, BF$_3$, CF$_2$, C$_2$H$_2$O$_2$, C$_4$H$_8$, CH$_4$-Ne, He-Ne, NH$_3$-H$_2$O, HF-HF, Ne$_2$, NH$_3$-NH$_3$, CH$_4$-CH$_4$, H$_2$O-H$_2$O;

\textbf{Mixed first- and second-row molecules:}
CS, OS, CS$_2$, PF$_3$, O$_2$S, SiO, CH$_3$S, AlF$_3$, F$_3$Cl, NOCl, HOCl, SiF$_4$, BCl$_3$, CH$_3$Cl, CH$_4$S, CHCl$_3$, COS, FCl, OCl, C$_2$H$_4$S, C$_2$Cl$_4$, ClCN, SiF, CH$_3$Cl-HCl, He-Ar, Ne-Ar;

\textbf{Second-row molecules:}
P$_2$, S$_2$, SH, H$_2$S, Cl$_2$, HCl, PH$_3$, SiH$_3$, SiH$_4$, AlCl$_3$, SiCl$_4$, SiH$_2$, AlH, AlH$_3$, HCl-HCl, H$_2$S-H$_2$S, HCl-H$_2$S.

In addition we considered 28 charged systems: 

\textbf{Singly-charged:} (H$_2$O)$^+$, (N$_2$)$^+$, (O$_2$)$^-$, (OH)$^+$, (C$_2$H$_2$)$^+$, (CH$_4$)$^+$, (CO)$^+$, (HCl)$^+$, (O$_2$)$^+$, (C$_2$H$_4$)$^+$, (HF)$^+$, (NH$_3$)$^+$, (OH)$^-$, (S$_2$)$^+$, (SiH$_4$)$^+$, (Cl$_2$)$^+$, (SH)$^-$, (S$_2$)$^-$, (SH)$^+$, (Cl$_2$)$^-$;

\textbf{Double cations:}
(C$_2$H$_4$)$^{2+}$, (C$_2$H$_6$)$^{2+}$, (C$_4$H$_4$)$^{2+}$, (CH$_2$O)$^{2+}$, (H$_2$S)$^{2+}$, (N$_2$H$_2$)$^{2+}$, (NH$_3$)$^{2+}$, (PH$_3$)$^{2+}$ .

The geometries of the investigated systems were taken from Refs. \cite{g2,w4,ae6,wi7,goerigk10,s22}.

To assess the performance of the ACBE method we compared its results with those of other popular extrapolation approaches. Specifically, the HELG \cite{helgaker97}, MART \cite{martin96}, and VARA \cite{varandas07} methods which are described in the Introduction. For first-row molecules we also considered the BAKO method \cite{bakowies07_2} (which is not defined for second-row elements). Moreover, in the case of DT-extrapolation, we also included the TRUH method \cite{truhlar98} (which is parametrized only for DT).

Because all these methods were originally developed and optimized neglecting core-valence correlation contributions and focusing mostly on first-row species, we have also performed a reoptimization of the parameters on our set of systems (including all the isolated atoms), in order to test further the capabilities of these approaches.
Therefore we have obtained the following three additional models:
POW-opt with $f(n)=n^{-\beta}$ and $\beta=1.804466$ for DT and $\beta=2.166051$ for TQ; OFF-opt with $f(n)=(n+l)^{-3}$ with $l=1.598529$ for DT and $l=1.331671$ for TQ;
2PAR-opt with $f(n) = (n+\alpha)^{-3}\left[1+\tau(n+\alpha)^{-2}\right]$ and $\alpha=1.739682$ , $\tau=0.894027$ for DT while $\alpha=1.452129$ , $\tau=0.943999$ for TQ.

The reference CBS values have been obtained, for each system, performing a three-parameter ($E_\infty$, $A$, $\beta$) fit based on Eq. (\ref{eq1}), with $f(n)=n^{-\beta}$, using TZ, QZ, and 5Z data.

\section{Results}
In this section, we compare the ACBE approach against several established extrapolation methods, evaluating its accuracy across neutral and charged species, first- and second-row atoms, and mixed systems.

\begin{table*}
%\begin{footnotesize}
\caption{\label{tab_neutral_mol_23}Statistics for the DT-extrapolation of neutral molecules: mean absolute error (MAE) in mHa, mean absolute relative error (MARE), and standard deviation in mHa.}
\begin{ruledtabular}
\begin{tabular}{lccccccccc}
%\hline\hline
 & HELG & TRUH & POW-opt & MART & OFF-opt & VARA & 2PAR-opt & BAKO & ACBE \\
 \hline
 \multicolumn{10}{c}{1st row molecules} \\
MAE	& 18.59 & 18.77 & 50.59 & 2.35 & 49.62 & 32.86 & 50.05 & 16.11 & 2.99 \\
MARE & 2.51\% & 2.76\% & 7.25\%	& 0.44\% & 7.11\% & 4.71\% & 7.17\% & 2.17\% & 0.46\% \\
Std.Dev. & 10.86 & 7.77 & 23.18 & 1.88 & 22.71 & 17.24 & 22.92 & 9.42 &	2.54 \\
 \multicolumn{10}{c}{Mixed 1st/2nd row molecules} \\
 MAE  &  170.17  &  76.66  &  23.66  &  118.05  &  23.81  &  38.73  &  23.74 & - &  29.46 \\ 
MARE  &  12.26\%  &  5.52\%  &  1.47\%  &  8.51\%  &  1.48\%  &  2.75\%  &  1.48\% & -   &  2.31\% \\ 
Std.Dev.  &  91.90  &  54.48  &  32.07  &  70.47  &  32.41  &  38.42  &  32.26  & - & 9.67  \\
 \multicolumn{10}{c}{2nd row molecules} \\
 MAE  &  206.48  &  119.36  &  45.17  &  157.93  &  47.43  &  79.33  &  46.42  &  -  &  12.53  \\ 
MARE  &  18.13\%  &  10.61\%  &  4.21\%  &  13.94\%  &  4.40\%  &  7.33\%  &  4.32\%  &  -  &  1.63\%  \\ 
Std.Dev.  &  135.04  &  76.81  &  28.06  &  102.55  &  29.49  &  46.50  &  28.85  &  -  &  10.72  \\
 \multicolumn{10}{c}{Overall} \\
 MAE  &  97.00  &  54.10  &  41.88  &  64.47  &  41.83  &  43.22  &  41.85  &  -  &  12.33  \\ 
MARE  &  8.21\%  &  5.02\%  &  5.03\%  &  5.27\%  &  5.00\%  &  4.64\%  &  5.01\%  &  -  &  1.20\%  \\ 
Std.Dev. &  113.55  &  73.02  &  45.75  &  90.43  &  46.35  &  54.66  &  46.08  &  -  &  13.51  \\
%\hline\hline
\end{tabular}
\end{ruledtabular}
%\end{footnotesize}
\end{table*}
\begin{table*}
%\begin{footnotesize}
%\begin{threeparttable}[b]
\caption{\label{tab_charged_mol_23}Statistics for the DT-extrapolation of charged molecules: mean absolute error (MAE) in mHa, mean absolute relative error (MARE), and standard deviation in mHa.}
\begin{ruledtabular}
\begin{tabular}{lccccccccc}
%\hline\hline
 & HELG & TRUH & POW-opt & MART & OFF-opt & VARA & 2PAR-opt & BAKO\footnotemark[1] & ACBE \\
 \hline
 \multicolumn{10}{c}{Singly charged} \\
MAE  &  75.90  &  47.20  &  32.68  &  54.75  &  33.12  &  37.11  &  32.92  &  8.97  &  4.42  \\ 
MARE  &  8.52\%  &  5.91\%  &  6.10\%  &  5.80\%  &  6.09\%  &  5.33\%  &  6.10\%  &  2.18\%  &  0.81\%  \\ 
Std.Dev.  &  89.97  &  59.33  &  34.44  &  72.81  &  35.16  &  43.73  &  34.84  &  4.51  &  6.54  \\
 \multicolumn{10}{c}{Double cations} \\
MAE & 35.97 & 26.59 & 33.09 & 23.21 & 32.89 & 27.75 & 32.98 & - & 9.00 \\
MARE & 6.29\% & 4.84\% & 6.49\% & 3.99\% & 6.44\% & 5.11\% & 6.46\% & - & 1.66\% \\
Std.Dev. & 42.84 & 34.24 & 28.59 & 37.85 & 28.73 & 30.51 & 28.67 & - & 8.44 \\
%\hline\hline
\end{tabular}
%\begin{tablenotes}
\footnotetext[1]{The statistic for the BAKO extrapolation method are only computed for the 1st row molecules, which are 12 over a total of 20.}
%\end{tablenotes}
%\end{threeparttable}
%\end{footnotesize}
\end{ruledtabular}
\end{table*}

The statistical analysis presented in Tables \ref{tab_neutral_mol_23} and \ref{tab_charged_mol_23} provides a comprehensive evaluation of the performance of various DT-extrapolation methods for both neutral and charged molecules. Across all categories, significant variations in accuracy are observed, with methods exhibiting different levels of mean absolute error (MAE), mean absolute relative error (MARE), and standard deviation.

For neutral molecules, the performance of the methods varies notably depending on the molecular subset considered. Standard methods (HELG, MART, TRUH, and VARA) generally perform well for 1st-row molecules, with MART achieving a particularly low MAE of 2.35 mHa. However, their accuracy deteriorates significantly for mixed 1st/2nd-row and 2nd-row molecules. This is especially evident for HELG, which reaches a MAE of 206.48 mHa for 2nd-row species, highlighting the inadequacy of the DZ basis set for these systems. This trend aligns with the fact that these standard methods were primarily designed for 1st-row molecules. Moreover, none of them were optimized for the aug-cc-pwCV$n$Z family of basis sets used in this study.

\begin{figure}
    \centering
    \includegraphics[width=0.9\columnwidth]{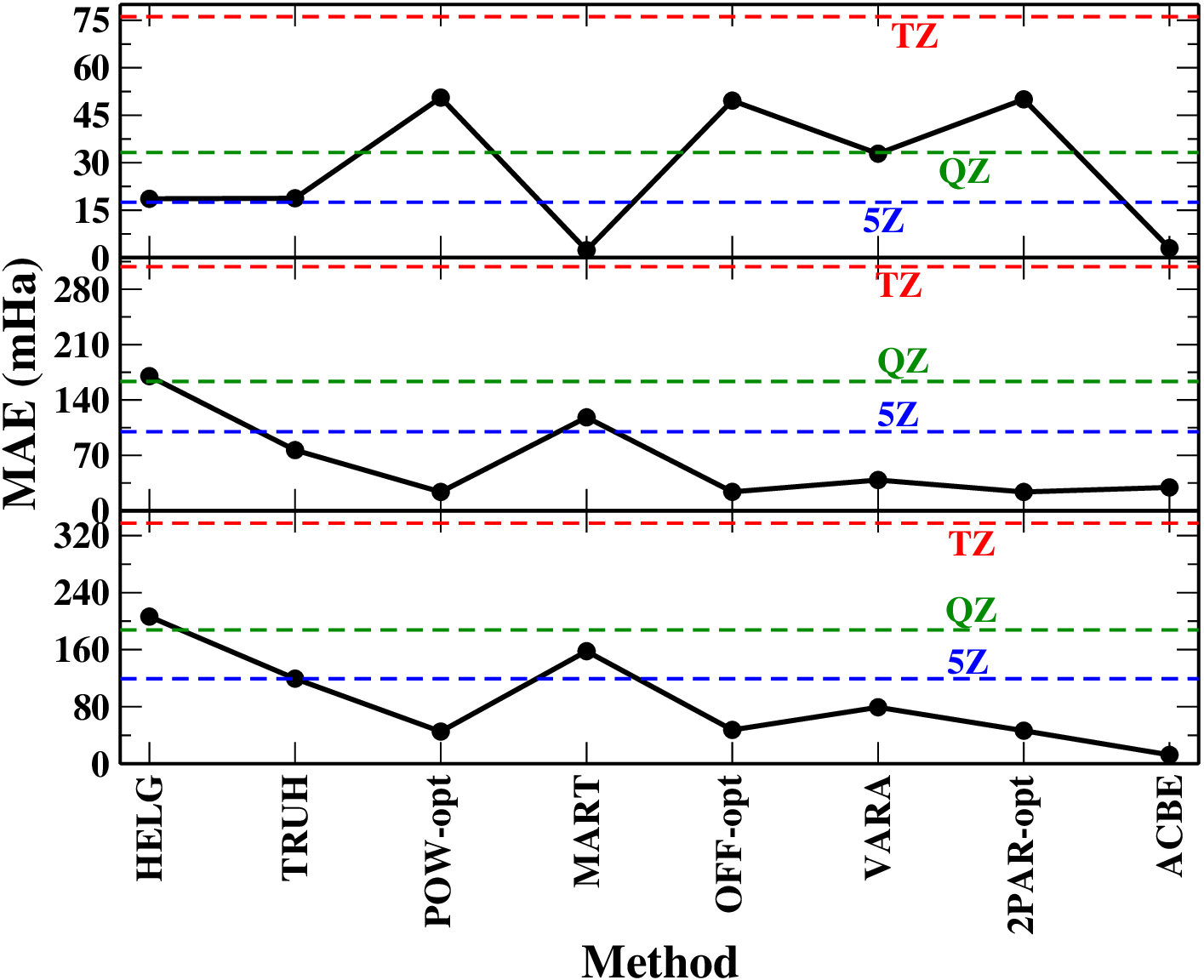}
    \caption{\label{fig_mae23}Mean absolute error (MAE) in millihartree (mHa) for different basis set extrapolation methods, categorized by molecular class (first-row, mixed first-/second-row, and second-row species). Results are obtained using double-zeta to triple-zeta (DT) extrapolation, illustrating the varying accuracy of each approach. Lower MAE values indicate better extrapolation performance.}
    
\end{figure}

In any case, as shown in Fig. \ref{fig_mae23}, all methods yield a considerable improvement over the performance of the TZ basis set. For first-row species in fact, the errors are comparable or lower than those obtained at the QZ level. For second-row elements, the advantage of using extrapolation methods is even greater, with most methods providing a performace comparable or superior to that of the 5Z basis set. 

\begin{figure}
    \centering
    \includegraphics[width=0.9\columnwidth]{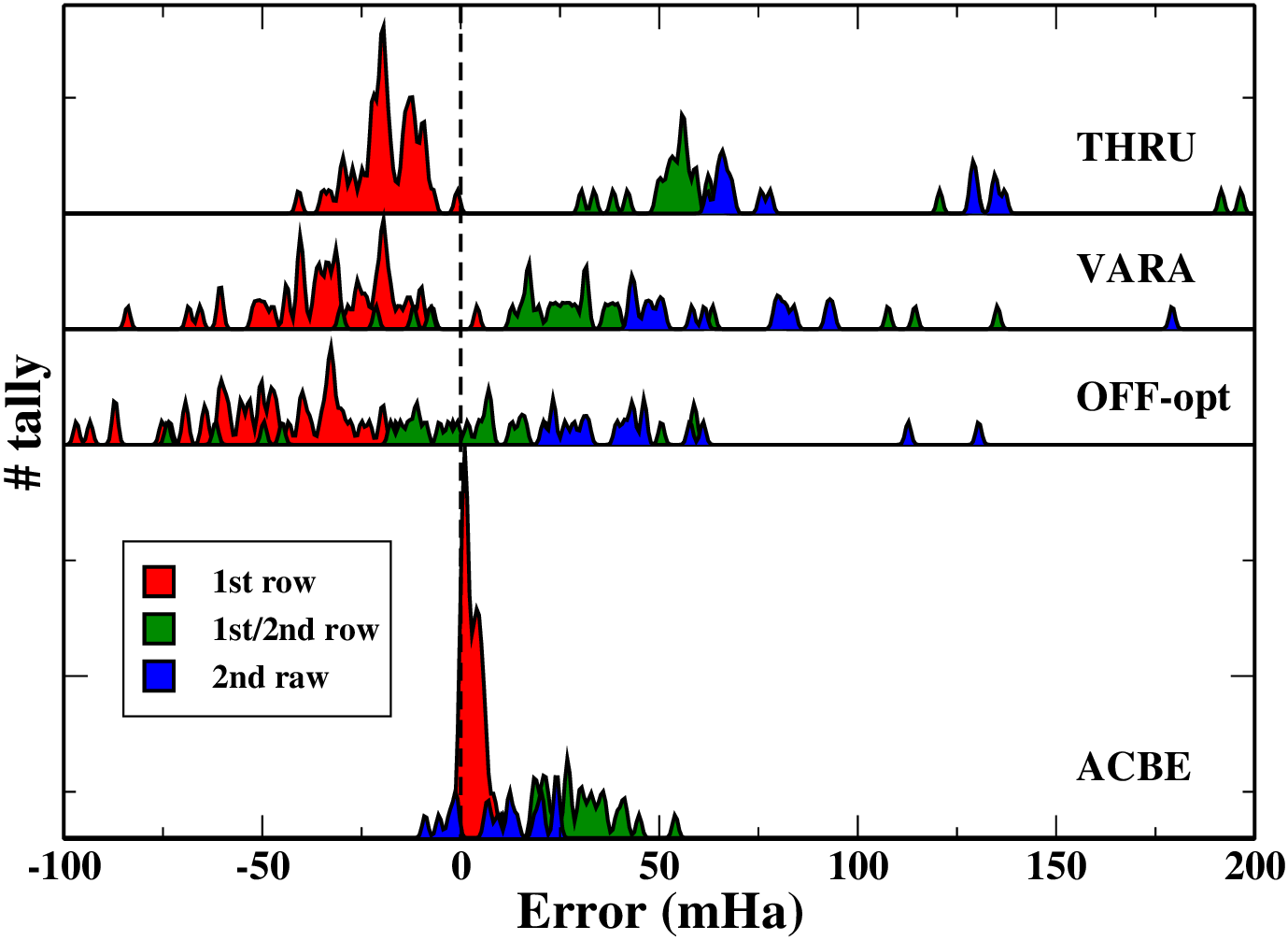}
    \caption{\label{fig_histo23}Distribution of extrapolation errors (in mHa) for selected DT-extrapolation methods across different molecular classes. This visualization highlights how error magnitudes vary among first-row, mixed first-/second-row, and second-row species.}
    
\end{figure}

To assess the limiting accuracy achievable, we also considered optimized counterparts of the standard methods, specifically tailored to the molecular set under study. The optimized methods (POW-opt, OFF-opt, and 2PAR-opt) exhibit a more uniform performance across different molecular subsets. For 1st-row molecules, they show larger errors than the standard methods, with MAE values exceeding 49 mHa. However, their accuracy improves for mixed 1st/2nd-row and 2nd-row molecules, yielding lower MAE values (approximately 23.7–23.8 mHa for mixed systems and 45.17–47.43 mHa for 2nd-row molecules). As illustrated in Fig. \ref{fig_histo23}, this behavior stems from the optimization procedure, which seeks a compromise in accuracy across molecular subsets with distinct characteristics. Consequently, despite being optimized for the dataset, these methods do not yield substantial overall improvements over their non-optimized counterparts.

This limitation arises from the use of a single global parameter, which fails to capture the diverse electronic environments within the dataset. Since no single parameter can accurately describe all species simultaneously, the improvements remain moderate at best, rather than consistent across all molecular subsets. This shortcoming motivated the development of a system-specific methodology, as implemented in the ACBE method. In fact, an inspection of the tables confirms that BAKO and ACBE deliver significant improvements with respect to other methods. Unlike the global parameter approaches, they employ system-specific parameters, allowing them to adapt effectively to the unique composition of each molecule.

However, ACBE, through Eq. (\ref{eq5}), also includes the specific adaptation of the extrapolation parameter to the molecular environment, which allows it to include the unique characteristics of each molecule. This feature, which is instead lacking in BAKO, helps to further improve the accuracy across all the molecular sets. In fact, if we neglect Eq. (\ref{eq5}) by setting all the $a_i$ and $b_i$ parameters to zero, the MAE of the modified-ACBE increases by 4 to 6 mHa for the different test sets. This is not a dramatic increase but shows the relevance of taking into account properly the electronic environment  and that the additional flexibility may result in considerably lower MAE and MARE values across all datasets.

An inspection of the tables confirms that ACBE delivers significant improvements in all cases. For neutral molecules, ACBE consistently outperforms the optimized methods, achieving an overall MAE of 12.33 mHa and a remarkably low MARE of 1.20\%. Notably, ACBE is also the only method that provides reliable accuracy for charged molecules, with an MAE of just 4.42 mHa and a MARE of 0.81\% for singly charged systems and 9.00 mHa (1.66\%) for double cations. On the contrary, all other methods exhibit significantly higher errors, with MAE values exceeding 30 mHa. Furthermore, ACBE maintains the lowest standard deviation among all methods, further demonstrating its robustness and consistency.

\begin{figure}
    \centering
    \includegraphics[width=0.9\columnwidth]{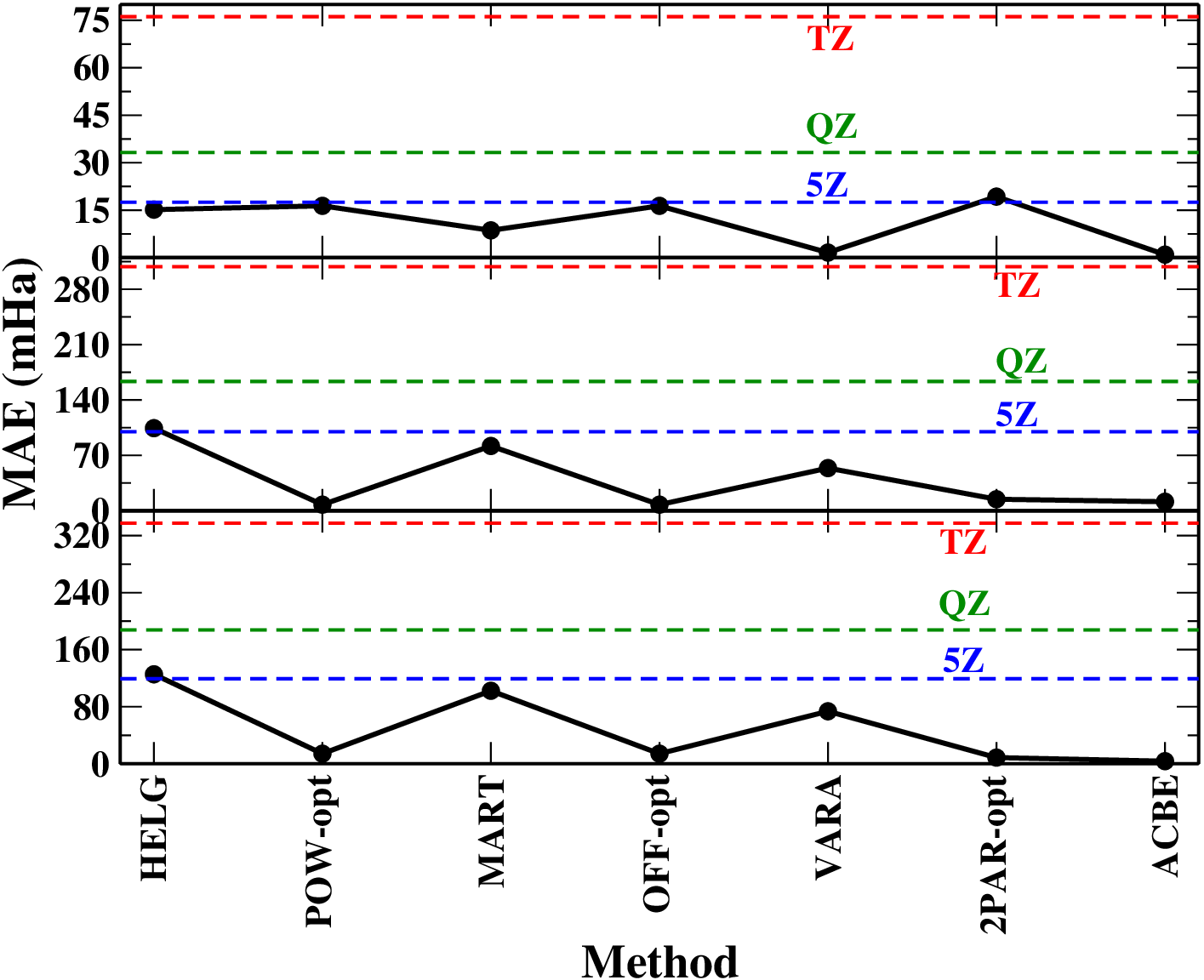}
    \caption{\label{fig_mae34}Mean absolute error (MAE) in millihartree (mHa) for different basis set extrapolation methods, categorized by molecular class (first-row, mixed first- second-row, and second-row species). Results are obtained using triple-zeta to qudruple-zeta (TQ) extrapolation, illustrating the varying accuracy of each approach. Lower MAE values indicate better extrapolation performance.}
\end{figure}

Turning our attention to the evaluation of TQ-extrapolation (Tables \ref{tab_neutral_mol_34} and \ref{tab_charged_mol_34}), we observe that the transition from DT- to TQ-extrapolation generally yields significant improvements in accuracy. This is reflected in the reduction of MAE, MARE, and standard deviation across all molecular subsets. As a result all methods are found to perform similar or often better than the 5Z level of theory (see Fig. \ref{fig_mae34}).
\begin{table*}
%\begin{footnotesize}
\caption{\label{tab_neutral_mol_34}Statistics for the TQ-extrapolation of neutral molecules: mean absolute error (MAE) in mHa, mean absolute relative error (MARE), and standard deviation in mHa.}
\begin{ruledtabular}
\begin{tabular}{lcccccccc}
%\hline\hline
 & HELG & POW-opt & MART & OFF-opt & VARA & 2PAR-opt & BAKO & ACBE \\
 \hline
 \multicolumn{9}{c}{1st row molecules} \\
MAE  &  15.17  &  16.38  &  8.63  &  16.38  &  1.60  &  19.28  &  3.45  &  0.94  \\ 
MARE  &  2.07\%  & 2.35\%  &  1.16\%  &  2.35\%  &  0.24\%  &  2.75\%  &  0.50\%  &  0.16\%  \\ 
Std.Dev.  &  8.74  &  7.25  &  5.59  &  7.25  &  2.05  &  8.65  &  1.77  &  1.13  \\
 \multicolumn{9}{c}{Mixed 1st/2nd row molecules} \\
 MAE  &  104.38  &  7.61  &  81.89  &  7.61  &  53.86  &  14.55  &  -  &  11.34  \\ 
MARE  &  7.52\%  &  0.57\%  &  5.90\%  &  0.57\%  &  3.88\%  &  1.07\%  &  -  &  0.86\%  \\ 
Std.Dev.  &  54.62  &  8.52  &  44.09  &  8.52  &  30.97  &  8.46  &  -  &  4.50  \\
 \multicolumn{9}{c}{2nd row molecules} \\
MAE  &  125.47  & 14.11  &  102.40  &  14.11  &  73.65  &  8.64  &  -  &  3.63  \\ 
MARE  &  11.17\%  &  1.58\%  &  9.18\%  &  1.58\%  &  6.71\%  &  1.16\%  &  -  &  0.47\%  \\ 
Std.Dev.  &  81.38  &  12.19  &  66.10  &  12.19  &  46.99  &  10.54  &  -  &  4.10  \\
 \multicolumn{9}{c}{Overall} \\
MAE  &  61.26  &  13.45  &  47.08  &  13.45  &  29.99  &  15.94   &  -  &  4.41  \\ 
MARE  &  5.33\%  &  1.70\%  &  4.01\%  &  1.70\%  &  2.49\%  &  1.98\%   &  -  &  0.42\%   \\ 
Std.Dev.  &  67.47  &  14.43  &  55.55  &  14.43  &  40.74  &  12.45  &  -  &  5.93     \\
%\hline\hline
\end{tabular}
\end{ruledtabular}
%\end{footnotesize}
\end{table*}
\begin{table*}
%\begin{footnotesize}
%\begin{threeparttable}[b]
\caption{\label{tab_charged_mol_34}Statistics for the TQ-extrapolation of charged molecules: mean absolute error (MAE) in mHa, mean absolute relative error (MARE), and standard deviation in mHa.}
\begin{ruledtabular}
\begin{tabular}{lcccccccccc}
%\hline\hline
 & HELG & POW-opt & MART & OFF-opt & VARA & 2PAR-opt & BAKO\footnotemark[1] & ACBE \\
 \hline
 \multicolumn{9}{c}{Singly charged} \\
MAE  &  46.34  &  8.88  &  36.17  &  8.88  &  24.12  &  8.44  &  5.61  &  1.29  \\ 
MARE  &  5.42\%  & 1.85\%  &  4.06\%  &  1.85\%  &  2.55\%  &  1.95\%  &  1.04\%  &  0.25\%  \\ 
STDEV  &  52.14  & 9.22  &  42.99  &  9.22  &  31.51  &  6.43  &  12.64  &  2.07  \\
 \multicolumn{9}{c}{Double cations} \\
MAE & 22.11 & 10.59 & 15.63 & 10.59	& 9.82 & 12.94 & - & 1.20 \\
MARE & 3.92\% & 2.10\% & 2.72\% & 2.10\% & 1.67\% & 2.56\% & - & 0.24\% \\
Std.Dev. & 23.80 & 8.17 & 20.23 & 8.17 & 15.85 & 7.11 & - & 1.38 \\
%\hline\hline
\end{tabular}
%\begin{tablenotes}
\footnotetext[1]{The statistic for the BAKO extrapolation method are only computed for the 1st row molecules, which are 12 over a total of 20.}
%\end{tablenotes}
%\end{threeparttable}
%\end{footnotesize}
\end{ruledtabular}
\end{table*}

For 1st-row molecules, TQ-extrapolation methods consistently achieve lower MAE and MARE values compared to their DT counterparts. A notable exception is MART, where the MAE increases from 2.35 mHa in DT to 8.63 mHa in TQ. This suggests that the strong performance of DT-extrapolation in this case was likely due to error compensation. However, other methods, such as VARA and ACBE, exhibit remarkable improvements, with MAE values decreasing from 32.86 mHa to 1.60 mHa and from 2.99 mHa to 0.94 mHa, respectively. The reduction in standard deviation follows a similar trend, confirming the enhanced stability of TQ-extrapolation approaches.

\begin{figure}
    \centering
    \includegraphics[width=0.9\columnwidth]{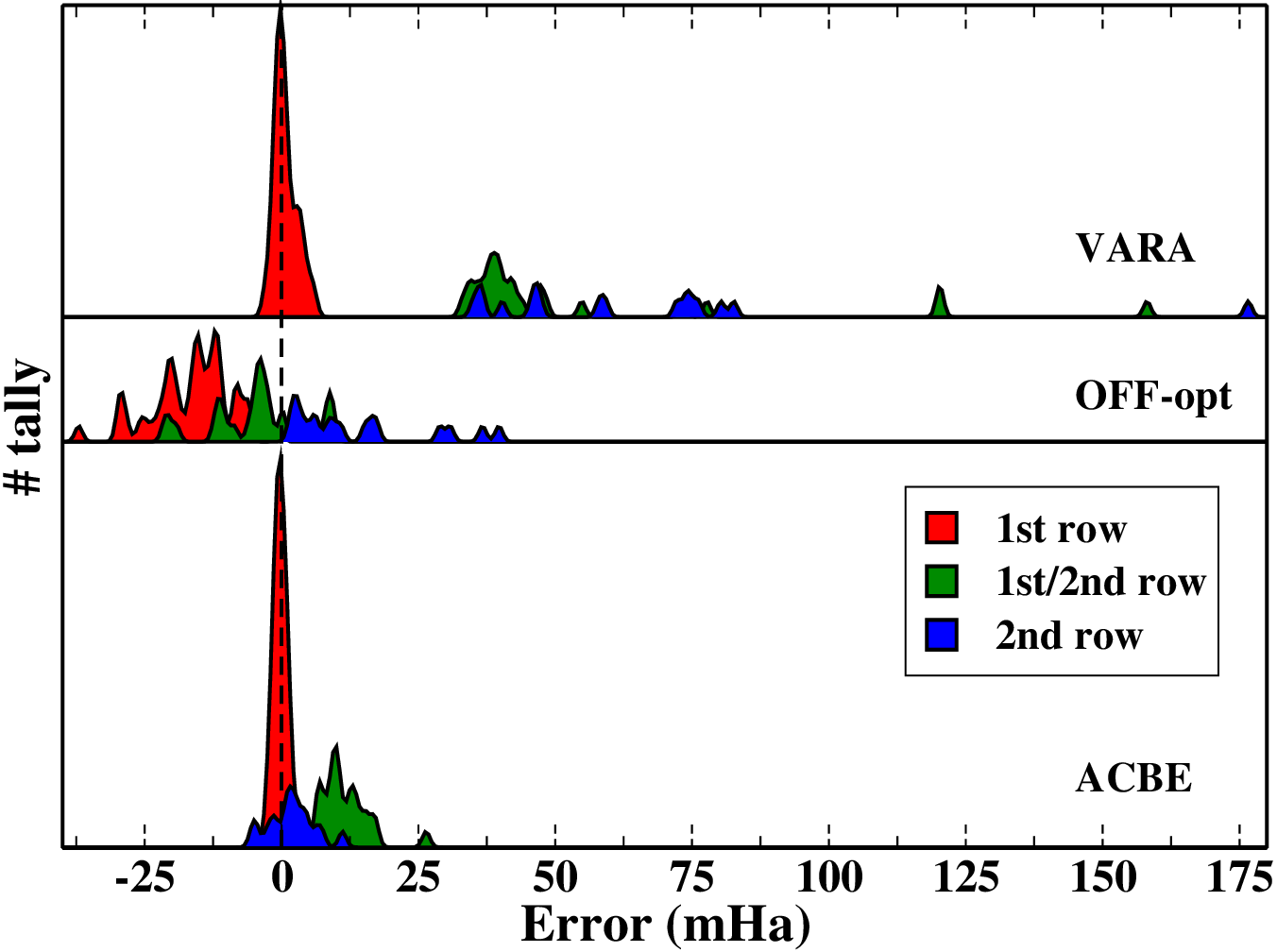}
    \caption{\label{fig_histo34}Distribution of extrapolation errors (in mHa) for selected TQ-extrapolation methods across different molecular classes. This visualization highlights how error magnitudes vary among first-row, mixed first-/second-row, and second-row species.}
\end{figure}

For mixed 1st/2nd-row molecules, the improvements are even more pronounced. Methods such as POW-opt and OFF-opt, which had MAE values of 23.66 and 23.81 mHa in DT, respectively, show significant reductions to 7.61 mHa in TQ. The ACBE method also demonstrates improved accuracy, with its MAE decreasing from 29.46 mHa in DT to 11.34 mHa in TQ. The standard deviations follow the same pattern, reinforcing the reliability of TQ-extrapolation methods across different molecular types. This is further illustrated in Fig. \ref{fig_histo34}.

For 2nd-row molecules, TQ-extrapolation significantly enhances accuracy. The MAE values for HELG and MART decrease from 206.48 and 157.93 mHa in DT to 125.47 and 102.40 mHa in TQ, respectively. Similarly, optimized methods such as POW-opt and 2PAR-opt show notable improvements, reducing their MAE from 45.17 and 46.42 mHa in DT to 14.11 and 8.64 mHa in TQ, respectively. ACBE continues to outperform all other methods, achieving a remarkable MAE reduction from 12.53 mHa in DT to just 3.63 mHa in TQ.

When considering overall performance, TQ-extrapolation provides a clear advantage over DT-extrapolation. The general MAE decreases from 41.88–97.00 mHa in DT to 13.45–61.26 mHa in TQ, depending on the method. Likewise, ACBE, which was already the best-performing approach in DT-extrapolation (12.33 mHa MAE), further improves in TQ-extrapolation, reducing its MAE to just 4.41 mHa. The overall MARE and standard deviation values exhibit similar trends, confirming the superior accuracy and stability of TQ-extrapolation.

\begin{figure}
    \centering
    \includegraphics[width=0.9\columnwidth]{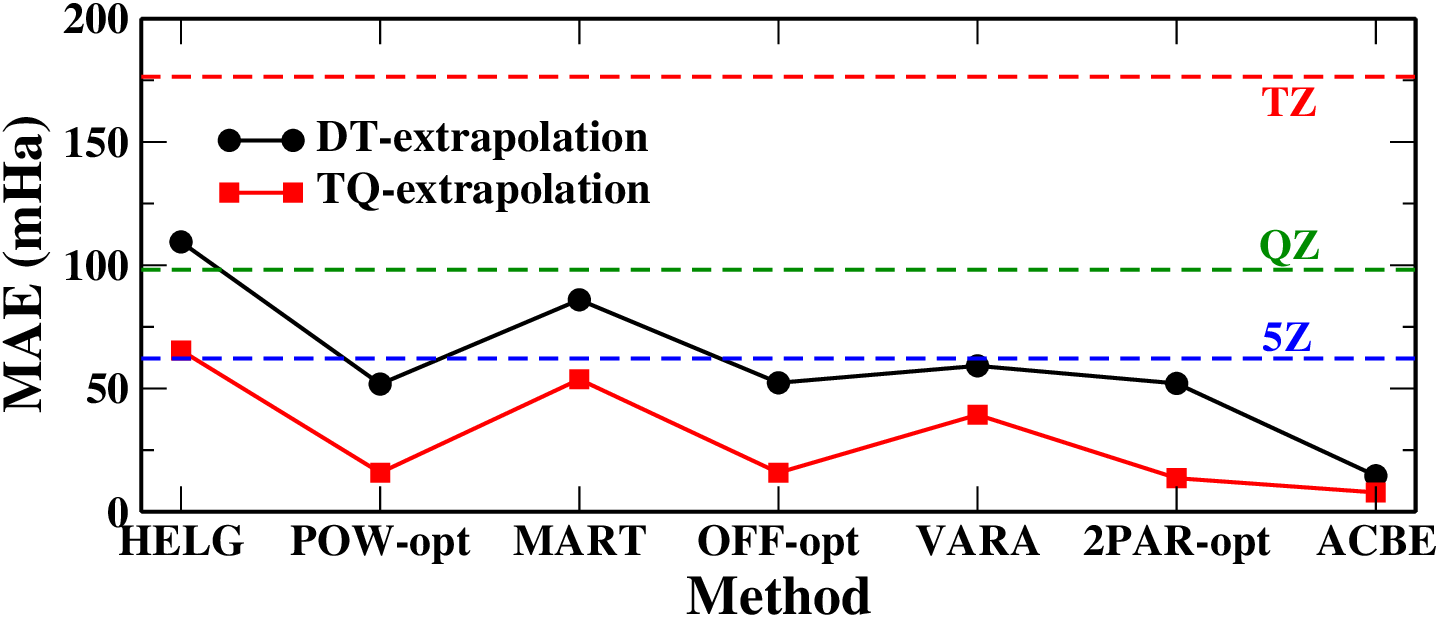}
    \caption{\label{fig_diff_mae}Mean absolute error (MAE, in mHa) for correlation energy differences computed using various extrapolation methods.}
\end{figure}

To complete our study, we analyzed MP2 correlation energy differences between the various systems, with the results summarized in Fig. \ref{fig_diff_mae}.
The observed trends closely follow those seen for absolute energies: DT-extrapolation methods achieve accuracy levels between QZ and 5Z, while TQ-extrapolation methods produce results that are comparable to or even surpass 5Z calculations. Notably, the ACBE method consistently delivers the best performance, yielding the lowest MAE across all cases.

Our analysis demonstrates that the ACBE method consistently delivers high accuracy and robustness. Compared to other widely used extrapolation techniques, it achieves low MAE and MARE values across all molecular subsets, including neutral, charged, first-row, second-row, and mixed-row species, excelling particularly in charged systems, where other methods tend to struggle.

These results establish ACBE as a computationally efficient and broadly applicable approach for improving MP2 extrapolation. By adapting to system-specific electronic environments, ACBE provides more accurate CBS energy estimates using small basis sets, achieving DT-extrapolation performance comparable to or even surpassing QZ-level calculations. This enables high-accuracy correlation energy predictions at a significantly reduced computational cost, making it a valuable tool for large and complex systems.

\section{Conclusions}
In this work, we have investigated the extrapolation of MP2 correlation energies using small correlation-consistent core-valence basis sets. We introduced and tested the Atom-Calibrated Basis-Set Extrapolation (ACBE) method, which incorporates system-specific parameters to improve accuracy and specificity for different systems. The ACBE method was systematically compared to a range of established extrapolation schemes, including both standard and optimized global parameter approaches.

By incorporating tailored parameters for each molecular system, ACBE circumvents the fundamental limitations of global approaches, which struggle to balance accuracy across diverse species, especially when small basis sets are used. Our results demonstrate that the ACBE method surpasses conventional extrapolation techniques, yielding the lowest MAEs across all tested molecular subsets, with a pronounced advantage for charged species, where global parameter-based methods fail to maintain accuracy. In particular, the ACBE approach exhibits strong performance in the DT-extrapolation regime, effectively capturing electron correlation effects with accuracy comparable to higher cardinal-number basis sets.

These results have significant implications not only for standalone MP2 calculations, where the ability to obtain accurate results via DT-extrapolation enables access to larger and more complex systems, but also for applications where MP2 energies serve as input quantities, such as double hybrids and, particularly, non-linear formulations such as adiabatic connection approaches. These advances may have broad importance for high-accuracy quantum chemistry, particularly in fields like dispersion interactions and high-rung correlation functionals.
Moreover, future studies could explore extending the ACBE methodology to different basis set families and beyond the MP2 level of theory, further enhancing its applicability and predictive power.

\section*{Supplementary material}
The supplementary material file reports all the MP2 correlation energies for each basis set and system considered in the present study.

\begin{acknowledgments}
We acknowledge the financial support from PRIN project no. 2022LM2K5X, Adiabatic Connection for Correlation in Crystals ($AC^3$) and 
ICSC - Centro Nazionale di Ricerca in High Performance
Computing, Big Data and Quantum Computing, funded by European Union - NextGenerationEU - PNRR.
\end{acknowledgments}

\section*{Data Availability Statement}
The data that support the findings of this study are available within the article and its supplementary material.

\nocite{*}
\bibliographystyle{unsrtnat}
\bibliography{main}

\end{document}